\documentclass[12pt]{article}
\usepackage{epsfig}
\begin{document}
\begin {center}
{\bf \Large Experiments needed in Meson and Baryon Spectroscopy}
\vskip 5mm
{D.\ V.\ Bugg\footnote{email: D.Bugg@rl.ac.uk} \\
{\normalsize\it Queen Mary, University of London, London E1\,4NS, UK}
\\[3mm]}
\end {center}
\date{\today}

\begin{abstract}
Three (or four) straightforward experiments would contribute
greatly to completing the spectroscopy of baryons and light
mesons.
In the baryon sector, data are needed on inelastic reactions
from a polarised target with $\pi ^\pm$ and $K^\pm$ beams up
to\,  $\sim 3$ GeV/c.
Similar data are needed in the light meson sector for
$\bar pp$ interactions in the momentum range 0.3 to 2 GeV/c.
In both cases, valuable information is to be obtained from
longitudinal (L) and sideways (S) target polarisations as well
as the conventional normal (N) polarisation.
Thirdly, $^3S_1$ and $^3D_1$ mesons in the mass range
1 to 2.4 GeV/c could probably be separated either by
diffractive dissociation of transversely polarised photons
or by $e^+e^-$ radiative return experiments
using transversely and longitudinally polarised electrons.

\vskip 2mm

\end{abstract}
\section {Introduction}
This is a discussion document aimed at stimulating discussion
of a fresh round of high quality experiments on baryon and
meson spectroscopy.
QCD is widely believed to lead to confinement of mesons
and baryons.
Lattice Gauge calculations are now possible with unquenched
quarks.
The opportunity exists to confront these calculations with
as complete a spectrum as possible of experimentally observed
states.
The experiments could be done quite cheaply and quickly.

This is not just `stamp-collecting', as some cynics
claim.
Without such data, it is not possible to say exactly how
QCD really works in the non-perturbative regime.
Understanding confinement is a key issue in particle
physics, but it is being neglected.
Confinement is clearly a phase transition, but it is quite
possible that it exhibits similar subtlety to chemistry and
solid state physics.
Chiral symmetry breaking is also clearly a phase transition.
Its relation to Confinement needs to be understood.
Perhaps they are the same phase transition, perhaps they
are related in a more subtle way.

Hybrids and glueballs are expected, but without a complete
picture of quark-model states, progress in identifying them
is frustrated.
What role, if any, do glueballs play in the confinement
process?
The baryon and light meson sectors are the ones where it is
presently technically feasible to achieve a complete
spectroscopy or something close.

Many $N^*$ and $\Delta$ resonances are known up to $\sim 2200$
MeV \cite {PDG}.
They are readily interpreted as 3-quark states.
However, the spectrum is incomplete.
Low spin states are missing, or poorly identified.
No member of the SU6 \{20\} multiplet is firmly
established, although there are candidates; perhaps they do
not exist.

Present baryon results come largely from experiments in the
1960-70 era using liquid hydrogen and polarised targets.
There are also good data on $\pi ^-p \to \eta n$
\cite {Nefkens} and a little on $\omega n$ \cite {Penner}.
The Crystal Ball collaboration has produced data on
$\pi ^- p \to \pi ^0 \pi ^0 n$ \cite {Prak1} and
$K ^- p \to \pi ^0 \pi ^0 \Lambda$ \cite {Prak2}.

Data on further inelastic channels come from bubble chamber
experiments but with low statistics; these have been analysed
by Manley and Saleski \cite {Manley}.
There are also low statistics data on $\Lambda K$ and $\Sigma
K$ final states, including polarisation information from hyperon
decays \cite {Saxon}.
Experiments on photoproduction are beginning to make decisive
contributions \cite {ELSA1, ELSA2, ELSA3, Thoma} but the photon spin
complicates the analysis.
The CLAS collaboration at JLAB proposes to take data
with polarised photons and a polarised target.
This will augment existing data \cite {Bradford}
and strengthen the partial wave analysis considerably.
Complementary information from $\pi N$, with its simpler
spin structure, would strengthen this partial wave analysis,
and  would also isolate couplings specific to
photons.

Because of the spin 1/2 of the nucleon, it is essential to
have polarisation data.
These data also fulfill a second important role.
Differential cross sections depend on intensities of partial
waves and the real parts of interferences between them.
The quantity $Pd\sigma /d\Omega$ depends on $\rm {Im}\, f^*g$,
where $f$ is the spin-averaged amplitude and $g$ the spin-flip
amplitude.
It is therefore phase sensitive and plays a key role in
establishing the phase variation of amplitudes.

%Fig. 1
\begin {figure} [htb]
\begin {center}
\label {Fig. 1}
\epsfig{file=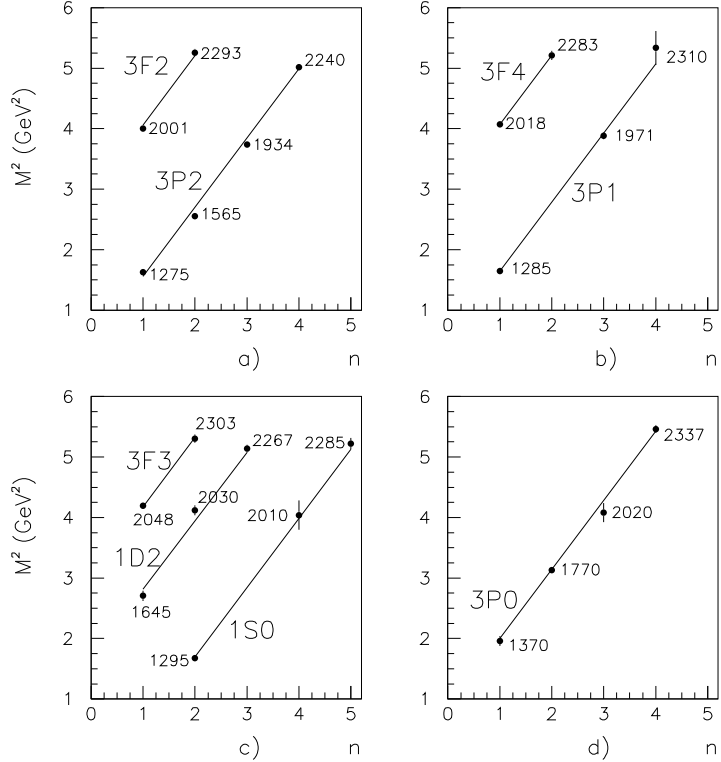,width=11.5cm}\
\vskip -6mm
\caption{Trajectories of $I = 0$, $C = +1$ light mesons:
$n$ is the Principal (radial) quantum number and n=1 for ground-states;
$M$ is the mass and points (with errors) show known states while
numbers indicate masses in MeV.}
\end {center}
\end {figure}

In the near future, several major $4\pi$ detectors will
complete their current programmes: Belle, Babar, Cleo C,
and Kloe.
They are superb detectors which have much to offer for
the experiments proposed here.
The one new feature which is required is a frozen spin
polarised target.
The technology of such targets is well developed and costs a
small fraction of the detectors themselves.
Rather than scrapping these detectors or cannibalising
them, why not put some of them to use on a new programme of
spectroscopy?
The statistics required are modest, so the required data
could be collected quite quickly.

The partial wave analysis would be the bigger problem.
Why not harness the efforts of the army of
phenomenologists who speculate on how QCD works to doing
the partial wave analysis and finding out how it really does
work?
This would be a welcome return to an earlier generation where
experimentalists and phenomenologists worked hand in hand.

In the light meson sector, the spectrum is incomplete around
1600-1700 MeV.
In the mass range 1910-2400 MeV, data from LEAR provide a
complete spectrum of $I=0$, $C = +1$ states, summarised
in the final coupled-channel analysis of  \cite {Anis1}.
This spectrum is currently not listed in the regular part
of the Particle Data Book, because this is the only
experiment to observe most of the states; results are to
be found on pp 644-648 of \cite {PDG}.
It is important to realise that many of the
states have been identified in as many as 7 independent
sets of data.
This makes identification of resonances extremely secure:
it can be shown by analysing sub-sets of these
data that the confidence level increases roughly as 2 to
the power of the number of data sets.
For $J^P = 0^+$, $2^+$ and $4^+$, the combined fit has a
confidence level better than the best individual set
by a factor 60.
This multiplicity of final states needs to be studied in
baryon spectroscopy.

Figs. 1 to 3 show the known light mesons above 1 GeV \cite {Bugg}.
They fall into a simple pattern of parallel Regge
trajectories.
Klempt has drawn attention to the fact that $N^*$ and
$\Delta$ states fall on similar trajectories of
almost the same slope, but with larger errors \cite {Klempt}.

An intriguing feature of both meson and baryon spectra
is the appearance of parity doublets: states with the same
isospin $I$ and spin $J$, but opposite parity $P$.
It is a feature of QCD that it is $SU(3)_L \otimes
SU(3)_R$ symmetric if quark masses are negligible.
However, it is well known that this symmetry is spontaneously
broken for the lowest states.
Firstly, the nucleon has no nearby $J^P = {1\over 2}^-$
partner.
Secondly, in the meson sector, well identified Adler zeros in
$\pi \pi$ and $K\pi$ elastic scattering arise from Chiral
Symmetry Breaking.

%Fig. 2
%\vskip -2mm
\begin {figure} [htp]
\begin {center}
\label {Fig. 2}
\epsfig{file=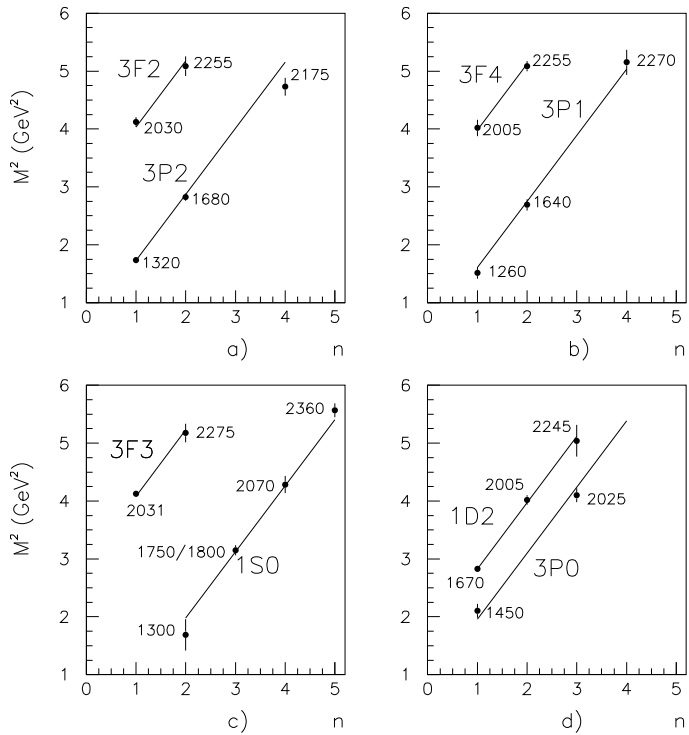,width=11cm}\
\vskip -4 mm
\caption{Trajectories of $I = 1$, $C = +1$ states
[16]}
\end {center}
\end {figure}

Glozman \cite {Glozman} proposes in a series of papers
that Chiral Symmetry is approximately restored high
in the spectrum.
This would require $J^{PC} = 4^{+-}$ mesons at masses
close to the well known $4^{++}$ states $f_4(2040)$
and $a_4(2040)$.
Likewise $3^+$ partners are needed for the well known
$3^-$ ground-states $\rho _3(1690)$ and $\omega _3(1670)$.
However, these states are currently missing \cite{Af3},
raising questions about this scheme.
Jaffe also comments on the question of
Chiral Symmetry restoration \cite {Jaffe}.
Shifman and Vainstein present disagreements with Glozman's
scheme \cite {Shifman}.

A feature of states labelled $^3P_2$ on Fig. 1 is that
they lie systematically lower in mass than those
labelled $^3F_2$ by $\sim 80$ MeV.
The labelling arises from the fact that $^3F_2$
states (a) decay mostly with $L = 3$ and
(b) are nearly degenerate with $^3F_3$ and $^3F_4$ mesons.
Likewise, $D$ states lie systematically above $P$ states by
$\sim 40$ MeV.
However, Glozman objects that orbital angular momentum should
not be a good quantum number for a rapidly rotating string
with highly relativistic quarks attached to each end.
Instead $J$ should be the good quantum number \cite {Gloz2}.

Afonin \cite {Afonin} provides an excellent general review
of the history and details of the spectroscopy.
He arrives at a different scheme where states fall more
naturally into hydrogen-like representations of the
dynamical O(4) group.
This difference from Glozman immediately illustrates the fact
that better and more complete data are required to settle even
the general features of how QCD actually works.
Afonin also traces a very interesting connection of
MacDowell symmetry from the baryon to the meson sector
\cite {Af2}.
Experiment might provide a useful guideline to
methods of approximation in Lattice Gauge calculations.

%Fig. 3
\vskip -6mm
\begin {figure} [htp]
\begin {center}
\label {Fig. 3}
\epsfig{file=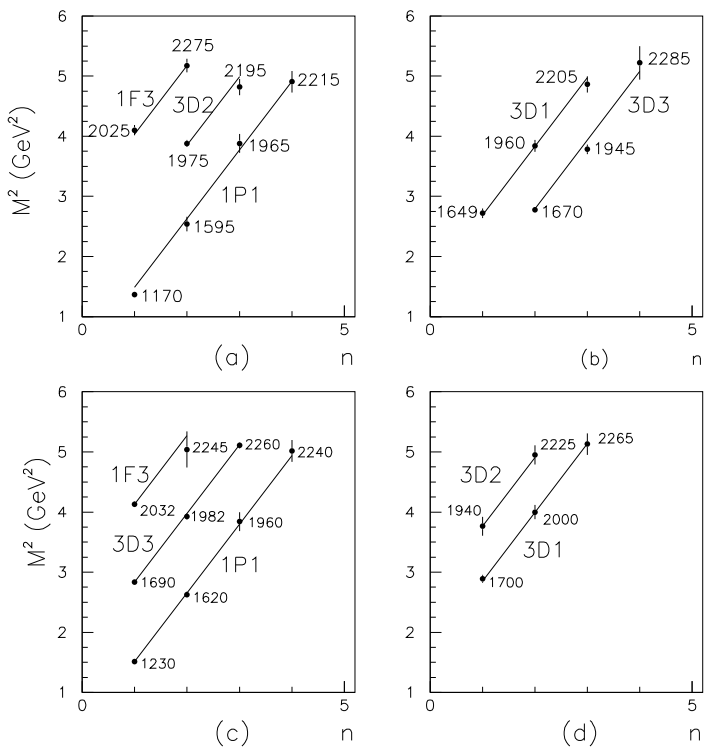,width=11cm}\
\vskip -4mm
\caption{Trajectories of $C = -1$ states with (a) and (b) $I = 0$,
(c) and (d) $I = 1$. In (b), the $^3D_3$ trajectory
is moved one place right for clarity; in (d), $^3D_2$ is moved one
place left [17,18]}
\end {center}
\end {figure}

The light $1^{--}$ mesons are particularly poorly
identified.
The essential reason for this is that each state
has a width $\ge 250$ MeV, but $^3S_1$ and $^3D_1$
states overlap at mass intervals less than this.
If Crystal Barrel data are a guide, $^3S_1$ states
should decay dominantly with $L = 0$ in the final
state (e.g. to $[b_1(1235)\pi]_{L=0}$) while $^3D_1$
states should decay dominantly with $L = 2$.
If so, an experiment on diffraction dissociation of
transversely polarised photons should make a clear
distinction between $^3S_1$ and $^3D_1$ states.
For $D$-wave decays, the partial wave decomposition is
\begin {equation}
|1,1> =
\sqrt {\frac {6}{10}}|2,2>_L|1,-1> -
\sqrt {\frac {3}{10}}|2,1>_L|1,0> +
\sqrt {\frac {3}{10}}|2,0>_L|1,1>.
\end {equation}
A linearly polarised photon is a superposition of initial
states $|1,1>$ and $|1,-1>$ with the result that
interference terms appear, generating distinctive dependence
on the azimuthal angle $\phi$ from the plane of polarisation,
through terms depending on $\phi$ and $2\phi$.

A linearly polarised photon beam of 9 GeV is planned for the
GlueX experiment at JLab \cite {Carman}.
It should be possible to adapt this beam for photons over
the range 1 to 2.5 GeV/c.
There is an excellent prospect that this would identify
individual $^3S_1$ and $^3D_1$ states cleanly.

The alternative (or a complementary experiment) is to
use so-called radiative return in $e^+e^-$ scattering using
polarised electrons. One hard photon is radiated from the
initial state and the surviving $e^+e^-$ pair interacts via a photon to
generate $J^P = 1^{--}$ final states.
It is desirable to use both linearly and longitudinally
polarised electrons, to distinguish clearly between
$L=2$ and $L=0$ decays of resonances.

So far, the discussion has centred on light mesons and
baryons.
There has, of course, been spectacular progress on
the spectroscopy of charmed mesons ($D$ and $D_s$-mesons) and
baryons and the family of states with hidden charm,
$J/\Psi$, etc.
B-factory data measure the spins of hyperon states from charmed
baryon decays; this has contributed strongly to identifying several
charmed baryons and also to identifying several $\Xi$ states.
Data from decays of $B$-mesons is contributing to the study of
charmed mesons, and also the study of light $q\bar q$ states and
hybrid candidates such as $Y_2(4260)$.

Lattice calculations are easier for heavier quarks.
If the spectroscopy of these states could be extended
to broad states and radial excitations, the spectrum would
be easier to compare with Lattice calculations.
But presently such data seem a remote prospect.
My essential message is that spectroscopy of light
mesons and baryons is a practical proposition over a
relatively short time-scale.
Perhaps, with this extra information, the spectroscopy of
charmed states would become clearer - at least one could
ask different questions.

\section {Practical Considerations}
In 1989, David Axen and I made a detailed study of a
possible experiment on  $\Lambda ^*$, $\Sigma ^*$
and $\Xi ^*$ spectroscopy \cite {Axen} using a polarised
target.
This paper is not available on-line; copies are available from the
authors. An outline of the contents of the paper will be given here.

It was written in the context of the $K$-factory being discussed at
the time.
It considered 15 readily accessible $K^-p$ reactions.
The study assumed a detector close to $4\pi$
acceptance and good $\gamma$ detection. It showed that it is realistic
to increase bubble chamber statistics by 2 to 3 orders of magnitude.

The design of the polarised target is simplified greatly if the
detector is used without a large-scale magnetic field.
The orientation of the target spin may then be manipulated simply
by means of holding coils of $\sim 0.25$ T.
A target length of 4 cm is realistic and a diameter $\le 1$ cm;
the holding coils then have dimensions $\sim 2$ cm larger than
the target in each direction.
The holding field bends charged particles by modest amounts,
but trajectories can be reconstructed iteratively.

Such a target operated at Triumf in the late 1980's and an operating
temperature of $0.06^\circ$K was achieved.
It may be possible to improve on that today.
Under those conditions, the relaxation time of the target polarisation
was typically 4 days.
The initial polarisation which was achieved was $85\%$.
The beam needs to enter along the cryostat of the polarised target.
This cryostat needs to be withdrawn from the detector from time to
time into a very uniform field of $2.5T$, where the target can
be polarised by the usual techniques.

\section {Baryon Spectroscopy}
Table 1 shows a selection of $\pi p$ reactions which should be
straightforward from a polarised target without a magnetic
field.
Column 3 shows the number of kinematic constraints
which are available if momenta are not measured.
It is assumed that photons are detected with the energy and position
resolution of typical CsI or BaF$_2$ detectors today.
The energies and directions of $\pi ^0$, $\eta$ and $\omega$ (from
$\pi ^0 \gamma$) are reconstructed fully.
The $\eta '$ may be detected via $\gamma \gamma$ or $\eta \pi ^0 \pi
^0$ with similar efficiency and resolution.

\begin {table}[htp]
\begin {center}
\label {Table 1}
\begin {tabular}{llll}
\hline
$\pi ^+p $ & $\to \pi ^+\pi ^0 p$ & 2C &(1) \\
           & $\to \pi ^+\eta  p$  & 2C &(2) \\
           & $\to \pi ^+\omega p$ & 2C &(3) \\
           & $\to \pi ^+\pi ^0 \pi ^0 p$ & 2C &(4) \\
           & $\to \Sigma ^+K^+$   & 3C &(5) \\
$\pi ^-p $ & $\to \pi ^-\pi ^0 p$ & 2C &(6) \\
           & $\to \eta  n$        & 3C &(7) \\
           & $\to \omega n$       & 3C &(8) \\
           & $\to \eta '  n$      & 3C &(9) \\
           & $\to \pi ^-\eta   p$ & 2C &(10) \\
           & $\to \pi ^-\omega p$ & 2C &(11) \\
           & $\to \pi ^0\pi ^0 \pi ^0 n$ & 3C &(12) \\
           & $\to \Lambda K^0$   &  2C & (13) \\
           & $\to \Sigma ^0K^0$   & 2C & (14) \\
           & $\to \Sigma ^-K^+$   & 2C & (15) \\\hline
$ \pi^+p$  & $\to \pi ^+\pi ^+ n$ & 2C &(16) \\
$ \pi^-p $ & $\to \pi ^+\pi ^+ \pi ^0 n$ & 2C &(17) \\
           & $\to \pi ^-\pi ^+ n$ & 2C &(18) \\
           & $\to \pi ^-\pi ^+ \pi ^0 n$ & 2C &(19) \\\hline
\end {tabular}
\caption {Easy channels for baryon spectroscopy; the upper
part of the table assumes particle momenta are unmeasured,
the lower part assumes neutron time-of-flight is measured.}
\end {center}
\end {table}

It is assumed that momenta of charged particles are not measured,
but $\pi$, $K$ and $p$ may be separated by $dE/dx$ and/or
time-of-flight.
In reactions (1)-(3), there are two unmeasured momenta, hence a 2C
fit after using energy-momentum balance for the production reaction.
Reaction (5) uses $\Sigma^+ \to \pi ^0p$.
There are 2 unmeasured momenta, but 4 constraints from energy-momentum
conservation in the production process plus a further constraint from
the relation between momentum and energy from the $\Sigma ^+$ decay;
$K^+$ identification makes this final state distinct.

Similar arguments give the number of constraints shown in the Table.
For reactions (7)-(15) it is assumed that neutrons convert in the
$4\pi$ detector, but without good time-of-flight information.
This situation could be improved using a dedicated time-of-flight
spectrometer at a distance of $\sim 2.5$m, covering 0-70$^\circ$
lab angles.
The event rate would drop accordingly, but beam intensity is not
a limitation and the trigger is simple.
For reactions (16)-(19) it is assumed that neutron time-of-flight
is measured.
There is a physics reason why it may be desirable to obtain limited
information on these channels, see Section 5.1 below.

With a $K^-$ beam of the same modest intensity, it is
possible to run a corresponding experiment on
strange baryons.
Here, a special trigger for all neutral final states
could be used to investigate rare channels like
$\Xi^0 K ^0$ and $\Xi^0 K ^0 \pi ^0$.
It would also be of interest to run a $K^+$ beam in case
there really are pentaquarks.

A Monte Carlo study of the background from
carbon or nitrogen atoms in the target estimates
background levels of $10\%$.
However, it is straightforward to measure this
background using a dummy target without hydrogen.
The separation of hydrogen events may be estimated roughly in the
following simple way.
Nucleons in carbon have a Fermi momentum of $\sim 200$
MeV/c, i.e. 115 MeV/c along each of $x$, $y$ and $z$-axes.
Momenta of final state particles need to be constructed
with an accuracy substantially better than this;
20 MeV/c is realistic.
For example the error for $\pi ^0$ reconstructed from photon
pairs, after using the constraint on the $\pi ^0$ mass,
is on average better than this.
Many of the reactions considered below have 2 constraints
from energy-momentum balance, so an elementary guess for
background levels is $3\%$; in practice, difficult
configurations make the backgrounds somewhat worse.

The conclusions of the 1989 Monte Carlo study were that:
\newline
1) reactions with integrated cross sections $\ge 1$ mb may
readily be separated by 2C fits with a level of cross-talk
between them and with other reactions generally in the
0.5-2.0$\%$ range,
\newline
2) the same is true for 3C fits to reactions with integrated
cross sections $\ge 10 \mu$b,
\newline
3) these levels of cross-talk are less than or equal to those
following from second scattering of exit particles
in the material of the polarised target and in the detectors, i.e.
intrinsic experimental limits.
\newline
4) Vertices of $\Sigma^\pm$, $\Xi$ and $K^0$ decays are identified
accurately by tracking the charged particles.

The Monte Carlo study is backed up by experience in
two experiments at Triumf \cite {Triumf} and Lampf
\cite {McNaughton} on the inelastic reaction
$pp \to pn\pi ^+$.
There, backgrounds averaging $8\%$ were observed,
and could be measured to $<1\%$.

Experience with Crystal Barrel is that statistics of
50K events per channel are adequate for partial wave analysis.
Suppose one aims for this with channels having an
integrated cross section of 1 mb.
This implies collecting a total of $\sim 6 \times 10^6$
events per momentum, allowing for the fact that roughly
half come from carbon (or nitrogen).
The Crystal Barrel experiment had a data acquisition rate of
60 events per second.
For more complex detectors, data acquisition rate may be
a problem, but the technology of LEP experiments is a
huge step forward.
Suppose an event rate of 100 events/s is possible.
Then 6 million $\pi p$ events can be recorded in $<24$
hours of running time.
With a 4 cm target length and a 30 mb average total
cross section, the required beam intensity is only 500
per second.
Using steps of 30 MeV/c from 500 to 3000 MeV/c,
the total running time for 80 momenta is a few months
per target spin orientation.
There is useful information from all three orientations
of target spin, so one is talking about a total of
1 year of data collection.

A technical point is that it is necessary to monitor
dead-times in the detector and data-acquisition system
carefully.
This is well known to be a systematic problem with most
large detectors.

Experience at the PS172 experiment at LEAR was that
differential cross sections can be deduced accurately
from data on a polarised target.
So it is not strictly necessary to run separately from a
liquid hydrogen target, though some systematic cross-checks
are desirable.
Some running from liquid hydrogen is needed for the
calibration of neutron detection efficiency.

\subsection {Three target polarisations}
It seems not to be realised widely that there is
useful information from target polarisations in
sideways (S) and longitudinal (L) directions.
For elastic scattering, where all the particles lie
in one plane, asymmetries from these spin orientations
are zero.
However, when the target spin is out of the plane of
the final state, asymmetries are non-zero.
These were measured in the Lampf experiment on
$pp \to pn\pi ^+$ \cite {McNaughton} and gave valuable
information.
Section 4 of that publication explains how to obtain
formulae for traces involving the Pauli matrices $\sigma$
representing target polarisation.
For hyperon decays in the final state, there is further
information from traces involving spin operators
for both target and hyperon.
The most useful data are for transverse target polarisation,
which gives information from real parts of interferences
between partial waves, but with signs different from
differential cross sections.
Longitudinal target polarisations measure
moduli squared of partial waves, but with signs which
depend on the sign of $L$.

\section {Light meson spectroscopy}
The Crystal Barrel experiment studied
16 final states containing only photons.
Background levels were as low as $1 \%$ for the strongest
channels, rising to $20\%$ in the worst cases.
However, there was so much physics information that the
cross-talk between channels could be estimated reliably.
Straightforward techniques were evolved to evaluate
cross-talk arising from 45 channels which were separated
at the stage of data-processing, although not all were
used for physics, see Section 2.2 of Ref. \cite {Bugg}.

For meson spectroscopy above 1900 MeV, the
essential idea is to study $s$-channel resonances:
\begin {equation}
\bar pp \to Resonance \to A + B.
\end {equation}
Channels which needs studying are shown in Table 2.
All of these final states are experimentally
easy to identify using decays of $\omega$ to $\pi ^0 \gamma$,
$\eta \to 2\gamma$ and $3\pi ^0$, $\eta ' \to 2\gamma$ and
$\eta \pi ^0 \pi ^0$.

%Table 2
\begin {table}[htp]
\begin {center}
\label {Table 2}
\begin {tabular}{llll} \hline $I=0$, $C =
+1$  & $I = 1$, $C = +1$ & $I = 0$, $C = -1$ & $I =1$,
 $C = -1$  \\\hline
$\pi ^0\pi ^0$, $\eta \eta$, $\eta \eta '$
& $\eta \pi ^0$, $\eta '\pi ^0$ & $\omega \eta$ &
$\omega \pi ^0$ \\
$\eta \pi ^0 \pi ^0$, $\eta '\pi ^0 \pi ^0$
 & $3\pi ^0$, $\eta \eta \pi ^0$ & $\omega \pi ^0 \pi ^0$ &
 $\omega \eta \pi ^0$ \\
\hline
\end {tabular}
\caption {Channels to be studied in $\bar pp$ interactions.}
\end {center}
\end {table}
%\vskip -4mm

It is of the greatest importance to obtain data at the lowest
available beam momenta.
The PS172 polarised target ran as low as 360 MeV/c $\equiv$
1910 MeV mass.
The intention was to run the Crystal Barrel experiment likewise
at 750 MeV/c and with 360 MeV/c at the target centre after energy loss.
However, LEAR closed before the allocated beam time was
delivered to the experiment, with the result that the
lowest momentum was 600 MeV/c $\equiv$ 1962 MeV.
Since there is a cluster of resonances in the mass range
1920 to 2050 MeV, this was a serious loss to the experiment.

For $I=0$, $C = +1$, the analysis is already very tightly
constrained for $J^P=0^+$, $2^+$ and $4^+$.
These partial waves appear in seven sets of data:
differential cross sections and polarisations in
$\bar pp \to \pi ^+\pi ^-$ \cite {Eisenhandler} \cite {Hasan},
and Crystal Barrel data on $\pi ^0 \pi ^0$, $\eta \eta$,
$\eta \eta '$, $f_2(1270)\eta$ and $a_2(1320)\pi$.
Phase information on other $J^P$ comes from their
interferences with these states in $\eta \pi ^0 \pi ^0$ and
$\eta ' \pi ^0 \pi ^0$ channels (and $\eta \eta \eta$).
Polarisation for these channels would make
the partial-wave analysis completely unambiguous.
Incidentally, the data and analysis programmes and fitting
parameters are publicly available on CDs from Sarantsev
or myself.
Partial waves are expressed in tensor algebra.
An important cross-check on programmes is that all partial
waves have been shown to be orthogonal when integrated over
phase space.

For $I=1, C = -1$, the polarisation information from
$\bar pp \to \pi ^+\pi ^-$ already makes the current
$q\bar q$ spectrum almost complete and rather well defined,
except for $J^{PC} = 1^{--}$, where overlap between
$^3S_1$ and $^3D_1$ states causes confusion.
For the other two sets of quantum numbers, the situation
is presently not nearly so well defined, because of the
lack of any polarisation data.
Results are consistent with spectra close to those of
the other two families, but several states are missing
and there is significant flexibility in the solutions.
Indeed two solutions exist for $2^+$ and $4^+$ partial
waves for $I=1$, $C=+1$, though one solution is significantly
better.
Simulations with current data show that polarisation
information on $\eta \pi$, $3\pi ^0$, $\omega \pi ^0$,
$\omega \eta$, $\omega \pi ^0 \pi ^0$ and $\omega \eta \pi ^0$
channels would make the analyses of $q\bar q$ states
completely unambiguous.

For the $3\pi ^0$ channel, there is the possibility that the
$\pi \pi$ $I=2$ channel could contribute.
[This is forbidden from initial $\bar pp$ states
with $I = 1$ and 0 in all other cases].
There is currently no evidence for $I=2$ $\pi \pi$, but that
needs confirmation.

This proposed experiment is not part of the PANDA programme at FAIR
\cite {Peters} which will run with $\bar p$ from 1.5 GeV/c
to 15 GeV/c on liquid hydrogen.
It would however be feasible at the lower momentum ring
FLAIR if there is physical space to accomodate a large
detector.
The existing Crystal Barrel detector is adequate for the task.
A beam intensity of a few $\times 10^4$ is needed with
a trigger on all neutral final states.
Other detectors have charged particle detection, which does
not survive in the present incarnation of the Crystal Barrel.
In reactions with charged pions in the final state, $G$ is
the relevant quantum number rather than $C$.
Accordingly, there are interferences between $I=1$ and $I=0$
states, providing very delicate cross-checks on parameters
of resonances in these two systems.

\section {Partial Wave Analysis}
There is a difference of opinion between experimental
groups as to whether to do so-called Energy Independent
Analysis or Energy Dependent Analysis.
In the former case, amplitudes are fitted freely in magnitude
and phase in every mass bin.
In the latter case, all partial waves are parametrised with
analytic forms.
In practice this means Breit-Wigner resonances plus
backgrounds linear in $s$, if needed.

The problem with the latter is that one is actively putting
resonances into the analysis, rather than `deducing' them
from the behaviour of amplitudes on the Argand diagram.
The converse problem with Energy Independent Analysis is that
it is theoretically impossible without polarisation
information.
In one bin, there is only one piece of information: a
differential cross section.
It is impossible in principle to derive both magnitudes and
phases of all amplitudes.
Energy Independent analysis therefore proceeds by
parametrising some well known resonances with Breit-Wigner
amplitudes, and using them as interferometers to determine
other partial waves.
This works well for the $\rho$ and $\pi _2 (1670)$, whose
parameters are well known.
However, for other resonances there are problems.
One example will suffice.
It is common practice to use the $f_2(1270)$ as an interferometer.
However, one usually finds with good data that the
$f_2(1565)$ is produced with it.
Its line-shape is not well known, because of lack of data
on the dominant $4\pi $ channel;
it overlaps significantly with $f_2(1270)$.
Worse, it is known to couple strongly to $\omega \omega$.
The sub-threshold continuation of this channel
can introduce serious distortions in what is fitted as
$f_2(1270)$.

The advantage of Energy Dependent analysis is that data can be
fitted simultaneously from a set of reactions.
In Energy Independent analysis, this constraint is lost unless
constraints from other data are imposed by fixing resonance mass,
width and/or branching ratios.
When a weak signal is analysed, this type of constraint is
important.
For the case of multi-body final states, e.g. $4\pi$, a combination
with analysis of 2-body channels is the only reliable approach.

Experience in both $\pi N$ and $\bar pp$ analyses is that the
constraint of analyticity plays a vital role, even when
polarisation data are available.
If this constraint is not applied, partial waves rapidly
deviate from analyticity.
One is instantly in a quick-sand of ambiguities.
Cross-talk develops between partial waves, making it
difficult to separate related partial waves, e.g.
$J^{PC} = 0^{++}$, $1^{--}$, $2^{++}$ etc,
and likewise $0^{-+}$, $1^{++}$, $2^{-+}$, etc.
Results which deviate significantly from analyticity cannot be
believed.
A further difficulty is that Energy Independent Analysis
requires literally thousands of parameters:
magnitudes and phases of all amplitudes in every bin.

The converse situation in Energy Dependent analysis
is that one must be careful to explore the maximum possible
variation of $s$-dependence.
For resonances, this is little problem in principle.
A Breit-Wigner amplitude of constant width is appropriate
unless one encounters thresholds.
If these thresholds are sharp, they are easily
accomodated by the Flatt\' e form
\begin {equation}
f = B(s)/[M^2 - s - i(g^2_1 \rho _1(s) + g^2_2 \rho _2(s))],
\end {equation}
for the two-channel case.
Here $g$ are coupling constants and $\rho$ are
phase-space factors $2k/\sqrt{s}$, where $k$ is momentum
in the decay channel.
Below threshold, $\rho $ needs to be continued analytically
or by means of a dispersion relation.
In the numerator, $B(s)$ is a centrifugal barrier factor needed
in most partial waves; form factors usually have negligible
effect over the widths of known resonances.

Complications arise however where the threshold opens
for production of a final state with significant
width, e.g. $f_0(1370) \to 4\pi$.
There is still a well defined prescription:
$\rho(s)$ needs to be integrated over the phase space of the
exit channel.
In addition, $M^2 - s$ needs modification to
$M^2 - s - m(s)$ where
\begin {equation}
m(s) =\frac {s - M^2}{\pi}\int \frac {ds'  M\Gamma
_{4\pi}(s')} {(s'-s)(s'-M^2)}.
\end {equation}
Evaluation of the dispersion integral needs to be done
only once if the form of $\Gamma_{4\pi}$ is known,
but the program then needs to interpolate in a table
of values; in practice this is trivial.

It is sometimes argued that `backgrounds' may be needed in
every partial wave.
This point needs clarification.
Left-hand cuts for each partial wave generate slowly varying
driving forces.
However, in the $N/D$ approach, these are isolated in
the $N$ function.
The denominator $D$ contains the phase information concerning
resonances.
It is not necessary to add separately left-hand cuts as
`backgrounds':
these are already built into resonances.
The classic example is Chew-Low theory \cite {Chew},
where nucleon exchange drives the $\Delta (1232)$
and alters its line-shape from a simple Breit-Wigner
by an amount which can in practice be used to make an
accurate determination of the $\pi NN$ coupling constant.

There may of course be broad components, e.g. hybrids or
molecular states, in addition to the regular $3q$ states,
and one needs to keep a watchful eye open for such
broad components.

The virtue of Energy Dependent Analysis is that
the number of fitted parameters is reduced to a few
per partial wave.
This has the advantage that programmes run quickly.
The downside is that one must be careful not to miss
something, requiring time to explore the options.
This procedure must be viewed as a process of successive
approximation.

The proof of the pudding is in the results.
Energy dependent analysis has dug out of Crystal
Barrel data a regular array of resonances.
Their star-rating can be investigated by varying
their parameters and dropping them completely from
the analysis to see what happens.
Analyses below the $\bar pp$ threshold have mostly
been done with Energy Independent Analysis.
The result, however, is a number of missing states
which can be confidently predicted from the quark
model, notably low spin states with $J^{PC} = 0^{-+}$,
$1^{++}$ and $2^{--}$.
My opinion is that these would emerge with judicious
use of analytic forms for amplitudes.
That needs to be done with existing data.

\subsection {Possible $I=2$ contributions}
There is one remaining issue which goes beyond current
partial wave analyses.
The $\pi \pi$ isospin 2 amplitude may contribute,
though there is presently no evidence for this.
If it does, its Clebsch-Gordan coefficients differ
between $\pi ^+p \to \pi ^+\pi ^+n$ and
$\pi ^+p \to \pi ^+\pi ^0p$ final states, for example.
To investigate this possibility, it is desirable to take some
data for the reactions in the bottom part of Table 1.
This can be done from a liquid hydrogen target using
$1C$ kinematic fits for these reactions.
The alternative is to use a time-of-flight spectrometer, giving
a 2C fit.

\section {Conclusions}
It is technically straightforward to use a
frozen-spin target in  a $4\pi$ detector to collect
data on baryon and light-meson spectroscopy.
Such a program could be completed in a few years and
would expand enormously the reliability and extent
of available data.
Data on $\pi N$ would strengthen the
results which can be deduced from the existing
photoproduction data (plus the measurements forseen
at ELSA with polarised target and polarised photons).

From such a program, it is predictable that the
spectroscopy of the regular baryons and mesons
could be determined completely up to 2400-2500 MeV,
i.e. over four radial excitations, which is surely
sufficient to see the picture.
Once this spectrum is established, the door is open
to uncovering glueballs with confidence in BES III
data on decays of $J/\Psi$, $\Psi '$ and $\Psi ''$.

It is very likely that the outlines of the hybrid
spectrum would also materialise.
In meson spectroscopy, there are presently three
good candidates: the well known $\pi _1(1600)$ with
exotic quantum numbers $J^{PC} = 1^{-+}$,
as well as two $2^{-+}$ states $\pi_2(1880)$ and
$\eta _2(1870)$, with masses  (and decay modes) which
do not fit regular $q\bar q$ states.
In addition the $\pi (1800)$ has decay modes
characteristic of those to be associated with a
hybrid; the problem here is that the $q\bar q$ state
expected at $\sim 1650$ MeV from the Regge trajectories
of Fig. 2(c) is as yet unknown and could be the $\pi (1800)$.

The cost of such a program in terms of new equipment is
small, and it would indeed extend the lives of existing
high quality detectors.
A significant effort of man-power is however required in
partial wave analysis.

\section {Acknowledgement}
I wish to thank Leonid Glozman, Sergey Afonin and Andrei
Sarantsev for extensive discussions of both theoretical
and experimental topics.

\end {document}